\documentclass[english,aps,prl,superscriptaddress,twocolumn]{revtex4-1}

\usepackage{natbib}
\usepackage{amsmath}
\usepackage{graphicx}
\usepackage{caption}
\usepackage{subcaption}
\usepackage{multirow}
\usepackage{hhline}
\begin{document}
\title{Neutron Limit on the Strongly-Coupled Chameleon Field}
\author{K. Li}
\affiliation{Department of Physics, Indiana University, Bloomington, Indiana 47408, USA}
\affiliation{Center for Exploration of Energy and Matter, Indiana University, Bloomington, IN 47408, USA}

\author{M. Arif}
\affiliation{National Institute of Standards and Technology, Gaithersburg, MD 20899, USA}

\author{D. G. Cory}
\affiliation{Department of Chemistry, University of Waterloo, Waterloo, Ontario N2L 3G1, Canada}
\affiliation{Perimeter Institute for Theoretical Physics, Waterloo, Ontario N2L 2Y5, Canada}
\affiliation{Institute for Quantum Computing, University of Waterloo, Waterloo, Ontario N2L 3G1, Canada}
\affiliation{Canadian Institute for Advanced Research, Toronto, Ontario M5G 1Z8, Canada}

\author{R. Haun}
\affiliation{Department of Physics, Tulane University, New Orleans, LA 70118, USA}

\author{B. Heacock}
\affiliation{Department of Physics, North Carolina State University, Raleigh, NC 27695, USA}
\affiliation{Triangle Universities Nuclear Laboratory, Durham, North Carolina 27708, USA}

\author{M. G. Huber}
\email[Corresponding author: ]{michael.huber@nist.gov}
\affiliation{National Institute of Standards and Technology, Gaithersburg, MD 20899, USA}

\author{J. Nsofini}
\affiliation{Department of Physics and Astronomy, University of Waterloo, Waterloo, Ontario N2L 3G1, Canada}
\affiliation{Institute for Quantum Computing, University of Waterloo, Waterloo, Ontario N2L 3G1, Canada}

\author{D. A. Pushin}
\email[Corresponding author: ]{dpushin@uwaterloo.ca}
\affiliation{Institute for Quantum Computing, University of Waterloo, Waterloo, Ontario N2L 3G1, Canada}
\affiliation{Department of Physics and Astronomy, University of Waterloo, Waterloo, Ontario N2L 3G1, Canada}

\author{P. Saggu}
\affiliation{Department of Chemistry, University of Waterloo, Waterloo, Ontario N2L 3G1, Canada}

\author{D. Sarenac}
\affiliation{Institute for Quantum Computing, University of Waterloo, Waterloo, Ontario N2L 3G1, Canada}
\affiliation{Department of Physics and Astronomy, University of Waterloo, Waterloo, Ontario N2L 3G1, Canada}

\author{C. B. Shahi}
\affiliation{Department of Physics, Tulane University, New Orleans, LA 70118, USA}

\author{V. Skavysh}
\affiliation{Department of Physics, North Carolina State University, Raleigh, NC 27695, USA}

\author{W. M. Snow}
\email[Corresponding author: ]{wsnow@indiana.edu}
\affiliation{Department of Physics, Indiana University, Bloomington, Indiana 47408, USA}
\affiliation{Center for Exploration of Energy and Matter, Indiana University, Bloomington, IN 47408, USA}

\author{A. R. Young}
\email[Corresponding author: ]{aryoung@ncsu.edu}
\affiliation{Department of Physics, North Carolina State University, Raleigh, NC 27695, USA}
\affiliation{Triangle Universities Nuclear Laboratory, Durham, North Carolina 27708, USA}

\collaboration{The INDEX Collaboration}

\date{\today}
\begin{abstract}
The physical origin of the dark energy that causes the accelerated expansion rate of the universe is one of the major open questions of cosmology. One set of theories postulates the existence of a self-interacting scalar field for dark energy coupling to matter. In the chameleon dark energy theory, this coupling induces a screening mechanism such that the field amplitude is nonzero in empty space but is greatly suppressed in regions of terrestrial matter density. However measurements performed under appropriate vacuum conditions can enable the chameleon field to appear in the apparatus, where it can be subjected to laboratory experiments. Here we report the most stringent upper bound on the free neutron-chameleon coupling in the strongly-coupled limit of the chameleon theory using neutron interferometric techniques. Our experiment sought the chameleon field through the relative phase shift it would induce along one of the neutron paths inside a perfect crystal neutron interferometer. The amplitude of the chameleon field was actively modulated by varying the millibar pressures inside a dual-chamber aluminum cell. We report a $95\,\%$ confidence level upper bound on the neutron-chameleon coupling $\beta$ ranging from $\beta <4.7 \times10^6$ for a Ratra-Peebles index of $n=1$  in the nonlinear scalar field potential to $\beta <2.4 \times10^7$ for $n=6$, one order of magnitude more sensitive than the most recent free neutron limit for intermediate $n$.  Similar experiments can explore the full parameter range for chameleon dark energy in the foreseeable future.  
\end{abstract}

\maketitle

\section{Introduction}

The discovery of the accelerated expansion of the universe~\cite{Riess98, Perlmutter99} in combination with other cosmological observations implies that a component of the universe called dark energy constitutes about $70\,\%$ of the energy density of the universe. This original work has been confirmed by more sensitive observations~\cite{Komatsu11, Larson11, Suzuki12, Sanchez12}. It is known that a contribution to the vacuum energy density acts like a negative pressure in Einstein's field equations, and since pressure gravitates it can cause the expansion of the universe to accelerate. There is no consensus on the nature and physical origin of dark energy, and most of the proposed research in this area consists of astronomical observations to more precisely characterize its effects on the universe's expansion rate and other cosmological and astronomical observables. However there is an interesting subset of ideas for the origin of dark energy which can be addressed in laboratory experiments. One set of such ideas postulate that dark energy is due to a scalar field $\phi$ which adopts a nonzero value in the vacuum of outer space.  For this scalar field to evolve into the dark energy seen today, one must postulate a self-interaction and dynamical screening mechanism to explain why it has not been observed in previous precision gravitational measurements. 


In this paper we specifically address a particular example of such a screened scalar field called the chameleon field~\cite{Khoury04a, Khoury04b, Brax04, Gubser04}. The chameleon field has a nonlinear potential of the form $V(\phi)= \Lambda^{4} + {\Lambda^{4+n} \over \phi^{n}}$~\cite{Ratra88} with $n$ the Ratra-Peebles index and $\Lambda$=2.4 meV based on the acceleration rate of the universe expansion. Additionally, it has an extra term that couples to matter field $ A(\phi)={\beta \over M_{PL}} \phi$, where $\beta$ is a dimensionless coupling to matter and $M_{PL}$ is the reduced Planck mass. So the effective potential takes the form $V_{eff}(\phi)=V(\phi) + A(\phi)\rho$. The appearance of the local matter density $\rho$ in the effective potential makes the effective mass $m^2_{eff} \sim n(n+1)\Lambda^{-{{4+n}\over {1+n}}}{({{\beta \rho} \over {nM_{PL}}})}^{{n+2}\over{n+1}}$ of the chameleon field density-dependent, allowing the chameleon to evade many of the existing experimental tests of gravity.  This also causes the chameleon field to be highly suppressed in the presence of even the modestly low matter density environment present in most terrestrial lab experiments where the effective mass is extremely heavy, thus further escaping detection.  As a result, the chameleon field, along with a range of similar theories, has yet to be ruled out by experiment.   A very extensive review of the chameleon field within the broader context of modified gravity theories has appeared recently~\cite{Joyce15}. 

This paper presents the result of a recent experiment using a perfect crystal neutron interferometer to place a limit on the chameleon field's coupling to matter $\beta$ and does so in a particularly direct, transparent manner. We actively modulate the amplitude of the chameleon scalar field in a gas cell in one arm of our interferometer and exploit the unique ability of neutrons to coherently penetrate the cell wall and access the phase shift from the neutron-chameleon coupling. This research activity brings together several physics subfields (gravitational physics, atomic physics, condensed matter physics) and in the neutron case employs centralized user facilities constructed mainly for materials science studies, thereby involving an uncommon diversity of scientific research techniques and environments in the quest to experimentally address what is perhaps the most exciting issue in cosmology. 


In the regime of small matter coupling $\beta$ the best laboratory constraints on chameleons over a wide range of $n$ come from laboratory tests of the inverse square law of gravity with sensitivity at the $100$ micron dark energy scale~\cite{Adelberger2007}. Experimental tests of the gravitational  inverse square law operating over other distance regimes, such as those designed originally to measure the Casimir interaction, can also constrain $\beta$. The sensitivity of a force sensor specially designed to search for chameleons has recently been analyzed~\cite{Almasi:2015}. 

In the regime of large $\beta$ techniques which employ two large masses start to become insensitive because both the source and sink of the chameleon field emanate only from a thin region of the surface of the objects due to the nonlinear chameleon self-shielding, and it is therefore preferable in this regime to employ test particles whose presence does not suppress the chameleon field. Neutron and atom interferometry can be used to search for the possible existence of chameleon and related scalar fields in this strongly-coupled chameleon regime since these probes do not  locally suppress the chameleon field. 

Slow neutrons can be used to perform sensitive searches for dark energy scalar fields~\cite{Pignol:2015}. The fact that neutron experiments are a sensitive method to probe chameleon fields was pointed out by~\cite{Brax11} in the context of an analysis of measurements on the quantum states of bouncing ultra-cold neutrons. The disturbance of the chameleon field near the surface of the flat neutron mirror employed in these measurements modifies the neutron bound state energies and wave functions as well as the relative phase of coherent superpositions of the neutron gravitational bound states. Recent experiments conducted in this system~\cite{Jenke:2014} have been used to constrain chameleon fields, and elaborations of this method are in principle capable of much greater sensitivity. Other neutron tests involving an apparatus used to test the weak equivalence principle for free neutrons~\cite{Pokot:2012} and a Lloyd's mirror type of neutron interferometer~\cite{Pokot:2013} have also been proposed.  A recent review of calculations performed to search for dark energy of various types using neutrons,  laboratory experiments to search for Casimir forces, and gravitational inverse square law violations has recently appeared~\cite{Brax:2015}.  

Calculations which showed that atoms can feel an unscreened chameleon field~\cite{Burrage:2014} have encouraged experiments to search for chameleons using atom interferometry. The first result of an atom interferometry experiment conducted to search for chameleons has appeared very recently~\cite{Hamilton:2015}. This experiment looked for a phase shift in a cesium atom interferometer operated in ultrahigh vacuum near a spherical mass which can be a source of a chameleon field. Already the constraints on the coupling $\beta$ from this atom interferometry experiment are quite strong. The prospects for further improvement in the atom interferometry experiments are very encouraging~\cite{Burrage:2015, Schlogel:2015}.

As the chameleon fields must couple directly to mass to be relevant for the observed universe expansion acceleration, other possible chameleon probes such as photons possess a more model-dependent coupling to these dark energy scalar fields. Strong experimental constraints on chameleon-photon couplings already exist. Examples of photon-based searches for chameleons include the CHASE (the GammeV CHameleon Afterglow SEarch) experiment~\cite{Chou:2009}, the Axion Dark Matter eXperiment (ADMX)~\cite{Rybka:2010}, a search for chameleon particles created via photon-chameleon oscillations within a magnetic field~\cite{Steffen:2010}, and the CAST Cern Axion Solar Telescope experiment~\cite{Anastassopoulos: 2015}.  

Most of the chameleon experiments performed to date using neutrons and atoms have sought the chameleon field by passing the probe close to a dense mass inside a high vacuum environment and searching for the phase shift from the chameleon-matter coupling $\beta$ in a chameleon field gradient. The chameleon field profile is obtained by solving the appropriate nonlinear Klein-Gordon equation for $\phi$ using the boundary conditions set by the experimental apparatus. A special feature of the experiment reported in this work is that we periodically introduce a nonzero matter density into the experimental chamber to actively suppresses the chameleon field in the measurement. To perform this type of search we exploit the fact that, unlike atoms, neutrons are able to pass through matter at densities (gas pressures of a few mbar suffice) for which the chameleon field is greatly suppressed. We were able to realize an experiment in which the chameleon field in the apparatus seen by the neutrons is repeatedly ``turned off" by the addition of a small gas pressure in the apparatus and ``turned on" by evacuation of the chamber. Moreover, using a neutron interferometer provides another advantage in that we can keep the pressure difference between the two arms of interferometer constant. This provides a direct measurement of any background that is independent of gas pressure, such as neutron scattering off the cell walls etc.



\section{Neutron Interferometry}

The experiment was performed at the National Institute of Standards and Technology's (NIST) Center for Neutron Research (NCNR) located in Gaithersburg, MD.  At the NCNR free neutrons are generated using a  20 MW research reactor which feeds over two dozen individual instruments that are primarily tailored for material science  applications.  Here we used monochromatic 11.1 meV neutrons and interferometric techniques similar to that of a Mach-Zehnder interferometer for light optics~\cite{Arif1993}. 

\begin{figure}[h]
\centering
    \includegraphics[width=\columnwidth]{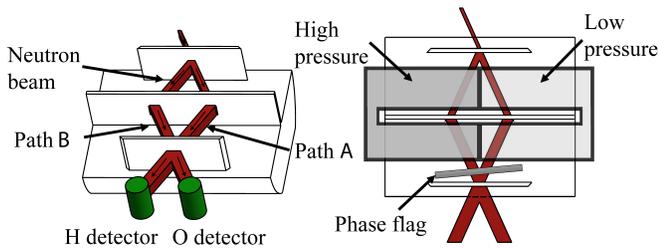}
    \caption{Left is a 3D schematic of the neutron interferometer seen in profile with the two coherent beam paths; right shows the top view of the two-chamber gas cell for the experiment, which fits around the central blade of the interferometer crystal.}
    \label{fig:interferometer}
\end{figure}

\begin{figure}[h]
\centering
\includegraphics[width=0.9\columnwidth]{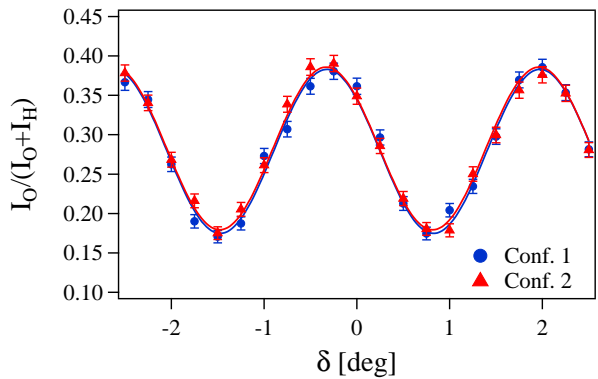}
\caption{A typical pair of O-beam interferograms, normalized to the total sum of counts, corresponding to the two run configurations in the experiment. Uncertainties are purely statistical.}
\label{fig:interferograms}
\end{figure}

The perfect crystal neutron interferometer used in this experiment consists of three crystal blades on a common crystal base; a schematic of which is shown in Fig.~\ref{fig:interferometer}. The monolithic silicon base below the blades ensures proper arcsecond alignment between the lattice planes of each of the three blades. The first blade serves to spatially separate the neutron's wave function $\psi e^{-i\Phi}$ into two coherent paths (A and B). In order for the two paths interfere, a central crystal blade  acts as a lossy mirror and directs the paths back together onto the third blade.  Neutrons exit the interferometer along either one of two paths labeled traditionally as  `O' and `H'  and are detected using highly efficient $^3$He-filled proportional counters.  It should be noted that there is only one neutron at a time inside the interferometer and thus it is a elegant example of macroscopic self interference.  Differences in phase $\Delta\Phi$ between the paths A and B modulates the intensities recorded by the detectors as
\begin{eqnarray}
\label{eqn1}
I_O = A_O + B\cos[\xi(\delta)+\Delta\Phi] \label{IO} \\
\label{eqn2}
I_H = A_H - B\cos[\xi(\delta)+\Delta\Phi] \label{IH}
\end{eqnarray}
\noindent 
In order to determine $\Delta\Phi$ and the other fit parameters ($ A_{O,H}$ and $B$) one could vary the cosine term in a controllable way ($\xi(\delta)$).  This is done by the adding what is called a  `phase flag' inside the interferometer.  The phase flag used here is a 1.5 mm thick $\times$ 50 mm wide piece of optically flat quartz and is illustrated in Fig.~\ref{fig:interferometer}b.  By rotating the phase flag an angle $\delta$ a phase shift of $\xi(\delta)$ is caused due to the effective path length difference  between paths A and B.  Rotating $\delta$ by $\pm$2.5 degrees creates an interferogram like the one shown in Fig \ref{fig:interferograms}.

The perfect crystal neutron interferometery technique employed in this work has been used to conduct a number of textbook experiments in gravitation, neutron optics, and quantum entanglement~\cite{RauchWerner}. These experiments include, but are not limited to, (1) the first demonstration that the gravitational field affects neutron wave functions as expected in non-relativistic quantum mechanics, (2) clear demonstrations of the fascinating minus sign in the quantum amplitude of a spin-$1/2$ particle rotated by $2\pi$, (3) the most precise determinations of neutron-nucleus scattering amplitudes, (4) sensitive tests of quantum entanglement predictions such as the Bell inequalities and the Greenberger-Horne-Zeilinger inequalities, and (5) subtle effects in neutron optics, most recently the successful manipulation of the orbital angular momentum quantum number of a neutron beam~\cite{Clark:2015}. 

Our experiment searches for the neutron phase shift between the two coherent paths of the interferometer arising from the coupling of the neutron to the chameleon field.  The neutron phase shift $\Phi_\mathrm{cham}$ due to the chameleon scalar field is
\begin{equation}
\label{eqn3}
\Phi_\mathrm{cham}= -\int{{\beta \over M_{PL}} {m^{2} \phi(x) \over {\hbar^{2}k}}dx}
\end{equation}
\noindent 
where $m$ is the mass of the neutron, $\phi(x)$ is the chameleon field, $k$ is the neutron wave vector, and the integration is performed over the neutron's trajectory. We use the sign convention that defines the neutron phase such that positive potentials give negative phases.  By measuring $\Phi_\mathrm{cham}$, we can then limit $\beta$ for a given Ratra-Peebles index $n$.

The calculation of the chameleon phase shift requires integrating over the neutron trajectory of the chameleon field profile $\phi(x)$ inside the cell as a function of the gas density. Previous calculations have shown~\cite{Brax:2013} that $\phi(x)$ is a rapidly varying function of the density of the gas. When the gas density is low the chameleon field $\phi(x)$ develops a nonzero amplitude for distances sufficiently far from the walls of the vacuum chamber. As the gas density is raised into a critical regime, which depends on $\beta$ and the geometry of cell, the chameleon field $\phi(x)$ is suppressed and tends toward zero. We can therefore actively modulate the chameleon field in the experiment. 

The first experiment to use perfect crystal neutron interferometry to search for chameleon scalar fields was recently performed at Institut Laue-Langevin (ILL)~\cite{Lemmel2015}. In this experiment neutrons also passed through a vacuum chamber mounted in the perfect crystal interferometer. The effect of the chameleon field was sought both by varying the relative separation of the neutron paths from the vacuum cell walls by translating the cell relative to the incident beam and also by varying the pressure inside the cell. Since the former is limited by the uniformity of a machined cell, in our experiment we employed the pressure variation method.

\section{Experimental Design and Setup}

A two-chamber vacuum cell is placed in the perfect crystal neutron interferometer (see Fig. \ref{fig:interferometer}b) with internal pressures controlled at different values. The vacuum cell as well as the two neutron beams are both symmetric about the center walls of vacuum cell, so those two neutron beams should feel exactly the same chameleon field if the pressure inside is same. However the neutron interferometer works in such way that it detects the difference between the two neutron beams so we need to control the pressure of the two chambers at different levels. First (Conf. 1), the pressure in the right chamber is kept low so that the chameleon field can develop a nonzero value. Meanwhile, the left chamber is filled with gas at a higher pressure so that the chameleon field is highly suppressed. Then (Conf. 2), the gas pressure in each chamber is raised by the same amount so that the chameleon field gets suppressed in both chambers. So the four phase shift neutron picks up during the experiment are,
\begin{eqnarray}
\Phi_\mathrm{1,A}&=&\Phi_\mathrm{cham,1A}+\Phi_\mathrm{cell,1A}+\Phi_\mathrm{gas,1A} \\
\Phi_\mathrm{1,B}&=&\Phi_\mathrm{cham,1B}+\Phi_\mathrm{cell,1B}+\Phi_\mathrm{gas,1B} \\
\Phi_\mathrm{2,A}&=&\Phi_\mathrm{cham,2A}+\Phi_\mathrm{cell,2A}+\Phi_\mathrm{gas,2A} \\
\Phi_\mathrm{2,B}&=&\Phi_\mathrm{cham,2B}+\Phi_\mathrm{cell,2B}+\Phi_\mathrm{gas,2B} 
\end{eqnarray} 
Ideally, the phase shift due to chameleon in $\Phi_\mathrm{1,B}$, $\Phi_\mathrm{2,A}$, $\Phi_\mathrm{2,B}$ should be close to zero so we define the phase shift of chameleon to be $\Delta\Phi_\mathrm{cham}=(\Phi_\mathrm{1,A}-\Phi_\mathrm{1,B})-(\Phi_\mathrm{2,A}-\Phi_\mathrm{2,B})$.

The helium gas pressures in the cell in either configuration are low enough that the equation of state of the helium gas is well-described by the ideal gas law. The gas density and resulting neutron phase shift from the neutron optical potential of the helium gas is then proportional to the gas pressure. At these low gas densities the neutron phase shift from the helium gas is a few orders of magnitude smaller than the ultimate sensitivity of our experiment to phase shifts from the chameleon field, so in practice $\Phi_\mathrm{gas}$ can be neglected. Furthermore, even if the phase shift from the gas was larger, our active control of the pressure difference between the two sides of the cell in the two configurations would cause this phase shift difference $(\Phi_\mathrm{gas,1A}-\Phi_\mathrm{gas,1B})-(\Phi_\mathrm{gas,2A} - \Phi_\mathrm{gas,2B})$ to cancel to high accuracy. The phase shift difference from the two cell walls in the two different configurations, $(\Phi_\mathrm{cell,1A}-\Phi_\mathrm{cell,1B})-(\Phi_\mathrm{cell,2A}-\Phi_\mathrm{cell,2B})$ is also negligible: the only physically plausible mechanism that might cause a difference, namely some absolute-pressure-dependent change of the phase shift from the neutron optical potential of the aluminum cell walls, is known to be negligible from previous measurements at much higher gas pressures at the NCNR of the neutron-helium optical potential. Under these conditions, $\Delta\Phi_\mathrm{cham}$ is a clean and direct measurement of the chameleon phase shift. Table \ref{table:pressure} shows the two sets of pressure configurations that are used in the experiment. The \lq\lq setpoint\rq\rq~in the table refers to a low enough pressure at which the chameleon field may produce an extra phase shift. The choice of gas pressures used in this experiment are also low enough that, for the chameleon coupling strengths to which we are sensitive, the chameleon field sees the helium gas used in the cell as a homogeneous medium based on previous analysis of this issue~\cite{Brax:2013}.

In the experiment we choose to control the absolute pressure (the low pressure side in Fig. \ref{fig:interferometer}) and the differential pressure across two chambers to eliminate possible systematic effects as discussed above. The absolute pressures in Conf.~2 will deviate slightly from the expected $1.33$ mbar and $2.67$ mbar shown in Table \ref{table:pressure} because the absolute pressure gauge used in this experiment has a maximum measuring range of 0 - $0.133$ mbar. So instead of measuring the pressures in Conf.~2 directly, the differential pressure gauge, which can measure up to $1.33$ mbar, is used to achieve the desired pressures in Conf.~2. We measured the associated pressure uncertainty to be smaller than $0.01$ mbar. This gives a negligible systematic uncertainty since the chameleon field amplitude is close to zero at such high pressure and it is only logarithmically sensitive to the pressure in this regime. 

\begin{figure}[h]
\centering
\includegraphics[width=0.9\columnwidth]{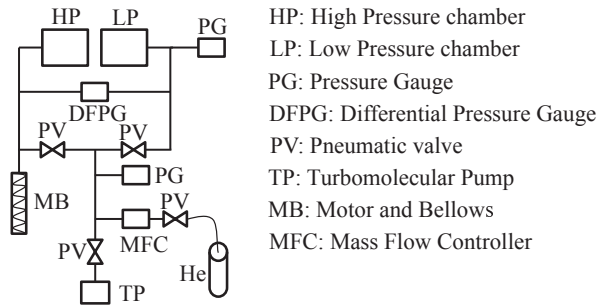}
\caption{Schematic diagram of the gas handling system which maintains a constant pressure in one chamber and a controlled pressure difference between the two chambers.}
\label{fig:ghs}
\end{figure}

Fig.~\ref{fig:ghs} shows the schematic of the gas handling system (GHS) used in this experiment. The vacuum cell is made of aluminum alloy 7075. Two CF flanges are machined on the two ends of the cell to accommodate aluminum gasket seals with negligible outgassing. A wall of thickness 3 mm separates the cell into two chambers which can be filled by gas at different pressures. The GHS employs metal seals (CF 1-1/3'') and ultra-high vacuum (UHV) compatible components which are helium leak tight as verified by measurements using a helium leak detector. All the vacuum tubes and the mechanical bellows have a high-conductance path to the cell (tubing diameters are greater than 2 cm) so the pressure gradient inside the gas handling system is minimized. Two absolute pressure capacitance transducers and one differential capacitance transducer are placed close to the vacuum cell and used to monitor the absolute pressure and differential pressure across the vacuum cell. Two types of feedback loops control the pressures inside the cells. One controls the differential pressure across the chamber using a motorized edge-welded stainless UHV bellows according to readings from the differential pressure gauge. The other feedback control employs an absolute pressure gauge and a vacuum compatible sensitive mass flow controller to control the pressure in the low pressure chamber. All valves are controlled by air-actuated switches to reduce possible electromagnetic or vibrational noise in the interferometer environment. Both absolute pressure and differential pressure are controlled with fractional fluctuations below $1\,\%$. Fig. \ref{fig:pressure} shows the pressure stability data.

\begin{table}[ht!]
\begin{center}
\begin{tabular}{ l | c | c | c}
  \hline			
  Run Cycle&Configurations & Pressure(High) & Pressure(Low) \\ 
  \hline
  \hline
  \multirow{2}{*}{Run 1} & Conf. 1 & 0.67 mbar & Setpoint \\
                        \hhline{~---}
                         & Conf. 2 & 0.79 mbar & 0.133 mbar \\ 
  \hline  
  \hline
  \multirow{2}{*}{Run 2} &  Conf. 1 & 1.33 mbar & Setpoint \\ 
                        \hhline{~---}
                          & Conf. 2 & 2.67 mbar & 1.33 mbar \\ 
  \hline  
\end{tabular}
\caption{Pressure configurations used for two different measurement runs.} 
\label{table:pressure}
\end{center}
\end{table}

\section{Calculation of the Chameleon Field Phase Shift in the Cell}

To calculate the chameleon scalar field inside a vacuum cell, one could solve the Klein-Gorden equation,

\begin{equation}
\Delta \phi = \frac{\partial V_{eff}}{\partial \phi}=\frac{-n\Lambda^4}{\phi^{n+1}}+\frac{\beta\rho}{M_{PI}}
\end{equation}

Unfortunately there is no known analytical solution to this non-linear second order partial differential equation so we had to use a finite difference method to obtain a numerical solution of the 3D field profile inside the experiential cell chamber. We built a 3D mesh with roughly $500\times500\times500$ nodes in total and solved the non-linear Poisson equation iteratively using the formula below (Eqn.~\ref{eqn:discretisation}, where $i$,$j$,$k$ denotes the node index in each dimension and $l$ denotes the iteration number). To accelerate the convergence of the calculation we further exploited the Gauss-Seidel method shown in Eqn.~\ref{eqn:iteration}. To get a precise solution, the number of grid points must be fairly large because the chameleon field grows rapidly close to the walls. The gradient of the chameleon field is close to infinity at such places, which inevitably causes instability in the iteration method. To address this problem, we implemented an uneven grid with more grid points where the field changes dramatically and fewer grid points where the field does not change much. The explicit formula is lengthy but is similar to Eqn.~\ref{eqn:iteration}.

\begin{widetext} 
\begin{equation}
\label{eqn:discretisation}
\frac{\phi^{(l)}_{i+1,j,k}+\phi^{(l)}_{i-1,j,k}-2\phi^{(l+1)}_{i,j,k}}{\Delta h_x^2}+\frac{\phi^{(l)}_{i,j+1,k}+\phi^{(l)}_{i,j-1,k}-2\phi^{(l+1)}_{i,j,k}}{\Delta h_y^2}+\frac{\phi^{(l)}_{i,j,k+1}+\phi^{(l)}_{i,j,k-1}-2\phi^{(l+1)}_{i,j,k}}{\Delta h_z^2} = \frac{-n\Lambda^4}{\phi_{i,j,k}^{(l)n+1}}+\frac{\beta\rho}{M_{PI}}
\end{equation}

\begin{equation}
\label{eqn:iteration}
\phi^{(l+1)}_{i,j,k}=\frac{(\frac{n\Lambda^4}{\phi_{i,j,k}^{(l)n+1}}-\frac{\beta\rho}{M_{PI}})+\frac{\phi^{(l)}_{i+1,j,k}+\phi^{(l+1)}_{i-1,j,k}}{\Delta h_x^2}+\frac{\phi^{(l)}_{i,j+1,k}+\phi^{(l+1)}_{i,j-1,k}}{\Delta h_y^2}+\frac{\phi^{(l)}_{i,j,k+1}+\phi^{(l+1)}_{i,j,k-1}}{\Delta h_z^2}}{\frac{2}{\Delta h_x^2}+\frac{2}{\Delta h_y^2} +\frac{2}{\Delta h_z^2}}
\end{equation}
\end{widetext}

\begin{figure}[h]
\centering
\includegraphics[width=0.9\columnwidth]{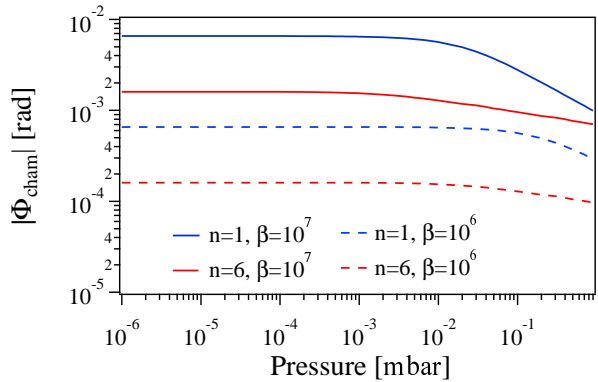}
\caption{The calculated phase shift caused by chameleon field versus gas pressure for various values of the Ratra-Peebles index $n$ and $\beta$.}
\label{fig:phase}
\end{figure}

\begin{figure}[ht]
\includegraphics[width=0.9\columnwidth]{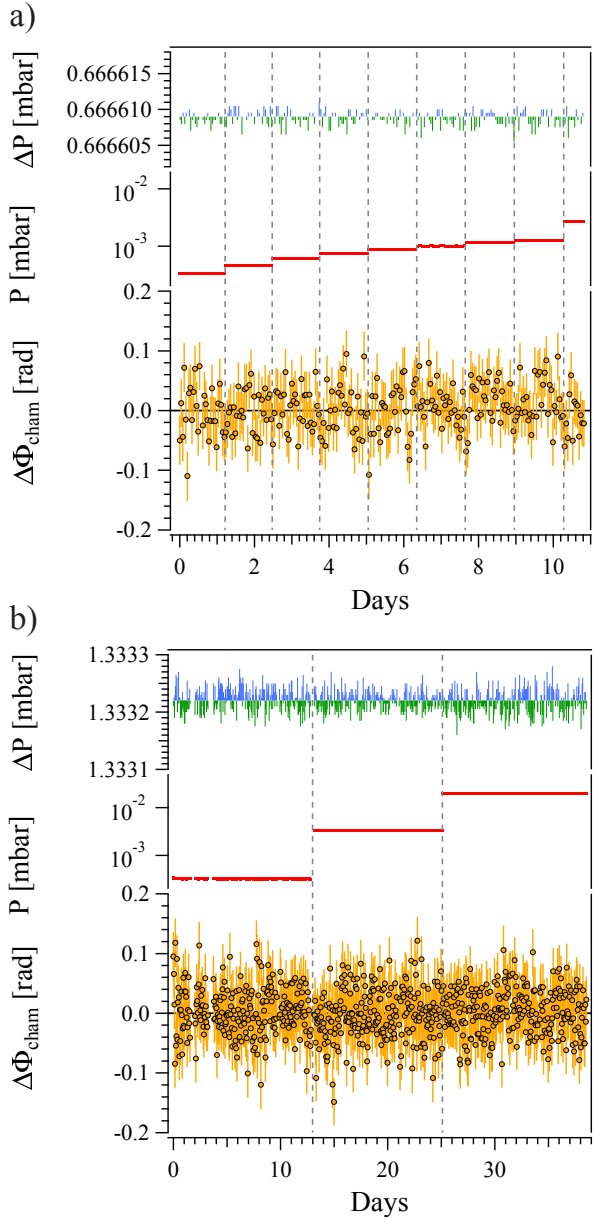}
\caption{a) and b) shows the pressure difference between the two cells ($\Delta P$), the Conf. 1 pressure in the low pressure chamber ($P$), and the measured chameleon phase shift in the first and second runs ($\Delta\Phi_\mathrm{cham}$).}
\label{fig:pressure}
\end{figure}

Having solved for the chameleon field inside vacuum cell, the extra phase shift picked up by neutrons due to chameleon field is computed according to Eqn.~\ref{eqn3} by integrating the chameleon field over the neutron path length. Fig.~\ref{fig:phase} shows the calculated phase shift caused by the chameleon field with different Ratra-Peebles model parameters $n$ and $\beta$. This result agrees well a previous calculation in the literature~\cite{Brax:2013}.

\section{Data acquisition and analysis}

Environmental factors are known to cause the phase of a neutron interferometer to drift.  To isolate the chameleon phase from environmental phase drifts, we switched between Conf. 1 and Conf. 2 after taking each interferogram. The contrast (or $B/A_O$ from Eqn.~\ref{eqn1}) of the neutron interference pattern with the cell in the interferometer was around $37\,\%$ and is consistent with other Al cells used inside the interferometer (the empty interferometer contrast is  $85\,\%$). Since the chameleon field is a function of pressure as shown in Fig. \ref{fig:phase}, we also varied the pressure setpoint in Table \ref{table:pressure} to look for any pressure dependence of the phase shift. Two sequences of measurements were taken during two adjacent reactor cycles. The only difference between the two runs is in the pressure configurations used. In the first run, nine pressure setpoints were scanned in the range from $3.33 \times 10^{-4}$ mbar to $2.67 \times 10^{-3}$ mbar. In the second run, we chose three pressure setpoints that span over a wider range, $3.33 \times 10^{-4}$ mbar, $3.33 \times 10^{-3}$ mbar and $2.00 \times 10^{-2}$ mbar. The measured phase shift along with the pressure and differential pressure stability is shown in Fig.5.

Systematic uncertainties in our measurement are negligible compared to the statistical uncertainty. Table~\ref{table:systematic} lists the major systematic uncertainties and corrections. The first two items could lead to a nonzero phase shift even in the absence of a chameleon field but both are much smaller than the statistical uncertainty (typically 0.0025 rad) in the chameleon phase. The last three factors reduce slightly the amplitude of the line integral used to compute the chameleon phase $\Phi_\mathrm{cham}$ (shown in Fig.~\ref{fig:phase}) and therefore reduces the calculated upper limit of $\beta$ proportionally. These scaling corrections are also negligible compared to the uncertainty in our upper limits for $\beta$. 

\begin{table}[h]
\centering
\begin{tabular}{|c|c|c|c|c|c|c|}
\hline
 systematic &  correction & uncertainty \\
 \hline
 Helium nuclear scattering & 0.002 rad/bar &  2.0E-6 rad \\  
 \hline
 Pressure gauge accuracy & $0.3\,\%$ FS & 1.2E-4 rad \\ 
 \hline 
 \multirow{2}{*}{Vacuum cell misalignment} & 1$^\circ$ rotation & $0.0005\Phi_\mathrm{cham}$ \\
 \hhline{~--}
 & 1 mm translation & $0.02\Phi_\mathrm{cham}$ \\
 \hline
 Neutron beam divergence & 1.5$^\circ$ & $0.006\Phi_\mathrm{cham}$ \\
 \hline
\end{tabular}
\caption{Estimates of systematic uncertainties.}
\label{table:systematic}
\end{table}

\begin{figure}[h]
\centering
\includegraphics[width=0.9\columnwidth]{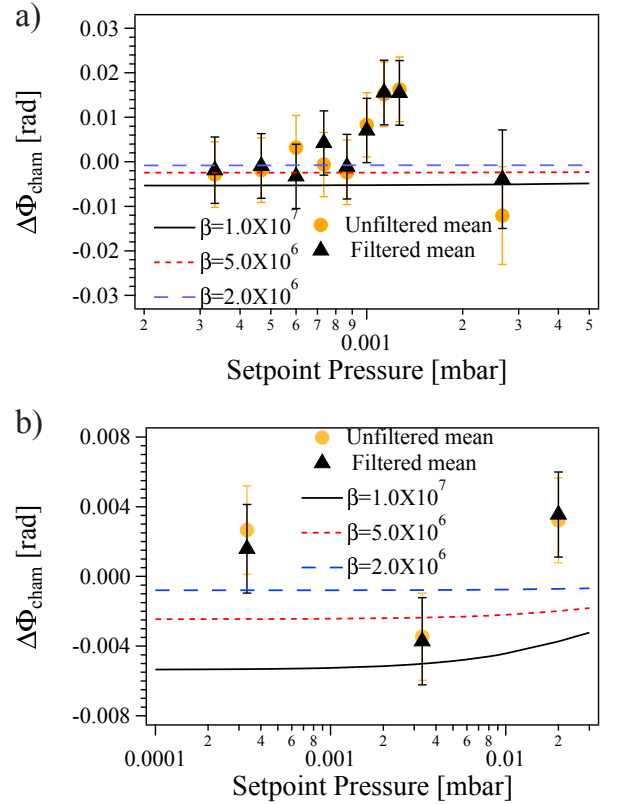}
\caption{a) and b) shows measured phase in two runs compared to calculated phase with
different values of $\beta$ (n=1).}
\label{fig:fit}
\end{figure}

Before fitting the measured phase shift to the calculated phase shift due to chameleon field, the sequence of phase shift differences is filtered using a time-series analysis algorithm designed to remove slow zero-point drifts in data sequences like ours which oscillate between two states with equal measurement times~\cite{Swanson}. To do this one takes a weighted average from neighboring points from the original sequence,  $y_i=\sum_{k=0}^{p}{c_ku_{i+k}}$, where $u_i$ is the original data sequence and $y_i$ is the combined sequence. A covariance matrix must be used to properly estimate the uncertainties after the correlations induced by the weighting algorithm. The weights $c_k$ satisfy the equation,

\begin{equation}
\begin{pmatrix}
  1 & 1 & 1 & 1 & \cdots & 1 \\
  0 & 1 & 2 & 3 & \cdots & p \\
  0 & 1 & 4 & 9 & \cdots & p^2 \\
  \vdots  & \vdots  & \vdots  & \vdots & \ddots & \vdots  \\
  0 & 1 & 2^{p-1} & 3^{p-1} &\cdots & p^{p-1} \\
  1 & 0 & 1 & 0 & \cdots & 0
 \end{pmatrix}
 \begin{pmatrix}
 c_0 \\
 c_1 \\
 c_2 \\
 \vdots \\
 c_{p-1} \\
 c_p
 \end{pmatrix}
 =
 \begin{pmatrix}
 0 \\
 0 \\
 0 \\
 \vdots \\
 0 \\
 1
 \end{pmatrix}
\end{equation}
which is designed so that a zero point drift in the signal with a polynomial time dependence up to order $p$ will be cancelled by combining each $p+1$ items in the sequence while a true signal correlated with the difference in the two configurations is kept unchanged. For comparison we present both the filtered mean and unfiltered mean in Fig. \ref{fig:fit}. The good agreement between the mean values and statistical uncertainties of the filtered and unfiltered phase shift data shows that any possible effects of interferometer phase drift are negligible in our measurement. We use the filtered data to extract our limit. 

\begin{figure}[h]
\centering
\includegraphics[width=0.9\columnwidth]{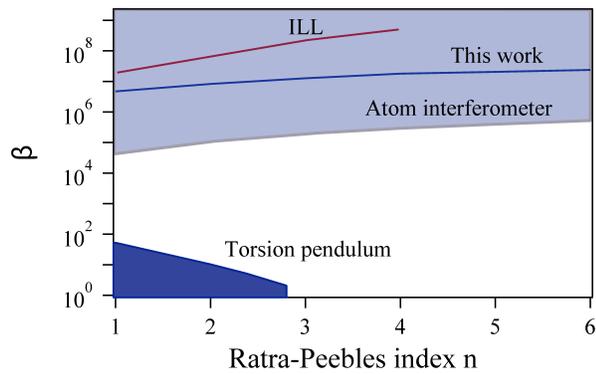}
\caption{The excluded regions in ($\beta$, $n$) parameter space compared to other experiments. From bottom to top: torsion pendulum experiment~\cite{Upadhye12a}; atom interferometer experiment~\cite{Hamilton:2015}; this work; ILL neutron interferometer experiment~\cite{Lemmel2015}. Other experimental constraints in the regime of large $\beta$ are less stringent.}
\label{fig:exclusion}
\end{figure}

To establish an upper limit of $\beta$ at the $95\,\%$ confidence level, the square of the weighted residuals is summed over all measurements ($\chi^2(\beta)$) 

\begin{equation}
\chi^2(\beta) = \sum_i {[\zeta(\beta)_i-\Delta\Phi_i]^2 \over \sigma_i^2}
\end{equation}

\noindent
where $\zeta(\beta)_i$ is the expected chameleon phase shift and $\Delta\Phi_i$ is the measured phase shift with uncertainty $\sigma_i$ for the $i^\mathrm{th}$ pressure setpoint.  To estimate $\beta$, $\chi^2(\beta)$ is then minimized with respect to $\beta$ for a given Ratra-Peebles index $n$.  However, the typical computation of a fit parameter confidence interval is not valid in this case, because for our measurements this function reaches its minimum at $\beta=0$ for all values of $n$ due to the constraint $\beta > 0$.  To find the 95\,\% confidence interval, $\chi^2(\beta)$ was solved for the value of $\chi^2$ that gives:

\begin{equation}
\int^{\chi^2 (\beta_\mathrm{lim} )}_0 p_{12} (\chi^{2 \prime}) d \chi^{2 \prime}  = 0.95
\label{eqn:chisqdist}
\end{equation}

\noindent
where $p_{12}(\chi^2)$ is the $\chi^2$ distribution with 12 degrees of freedom, corresponding to the 12 pressure set points.  The calculated limit is shown in Table \ref{table:result} and the excluded area is shown in Fig. \ref{fig:exclusion} as a function of the Ratra-Peebles index $n$.

\begin{table}[h]
\centering
\begin{tabular}{|c|c|c|c|c|c|c|}
\hline
 n &  1 & 2 & 3 & 4 & 5 & 6 \\
 \hline
 $\beta_{limit}\times10^6$ & 4.7 & 8.2 & 12.7 & 17.9 & 20.4 & 23.8 \\  
 \hline 
\end{tabular}
\caption{The calculated upper limit on $\beta$ with $95\%$ confidence level.}
\label{table:result}
\end{table}

\section{Conclusion}

We have conducted a search for chameleon dark energy fields using neutron interferometry. We realized an experiment in which the chameleon field is periodically varied in magnitude with no change in the experimental geometry.  Our upper bound of $\beta <4.7 \times10^6$ for a Ratra-Peebles index of $n=1$  to $\beta <2.4 \times10^7$ for $n=6$ is the most sensitive direct constraint on the free neutron-chameleon coupling in the strong coupling regime of the theory. It is more sensitive than a recent neutron interferometer experiment at the ILL~\cite{Lemmel2015} by about a factor of $5$ for $n=1$ to a factor of $30$ at $n=4$, and it cuts into part of the projected sensitivity of a proposed experiment using an optimized force sensor~\cite{Almasi:2015}. The constraints from this work on the chameleon are consistent with but less stringent than a very recently-published atom interferometry experiment using cesium atoms~\cite{Hamilton:2015}. Under the assumption that chameleon dark energy obeys the gravitational equivalence principle and that there are no essential differences between the response of a neutron and a cesium atom to the chameleon field, the atom interferometer constraints are about two orders of magnitude more stringent at present.    

This neutron interferometer experiment can be improved by (a) using an interferometer crystal with a larger path length, (b) improving the contrast of the interference signal in the interferometer, (c) optimizing further the pressure range of the measurements, (d) operating the interferometer on a more intense monochromatic neutron beam, and  (e) varying the neutron coupling to the chameleon both by changing the cell geometry and also by varying the gas pressure. With these improvements, the statistical sensitivity of this measurement to the coupling $\beta$ can be improved by at least two more orders of magnitude with negligible systematic effects, which could then surpass the existing atom interferometer limits at larger $n$. 

A experimental lower bound on $\beta>50$ for $n=1$ already exists from gravitational inverse square law tests. By improving the atom interferometry limits by another few orders of magnitude,  laboratory experiments will either discover chameleons or provide the first experimental refutation of a plausible dark energy theory. Scalar field candidates for dark energy which employ other screening methods, such as symmetrons, might also be constrained by this and other experiments with further analysis.

\section{Acknowledgements}

We acknowledge the support of the National Institute of Standards and Technology, US Department of Commerce, in providing the neutron facilities used in this work. This work was supported by NSF grants NSF PHY-1205342, PHY-1068712, PHY-1307426 and DOE award DE-FG02-97ER41042. Financial support provided by the NSERC CREATE and DISCOVERY programs, CERC, and the NIST Quantum Information Program are gratefully acknowledged. K. Li and W. M. Snow acknowledge the support of the Indiana University Center for Spacetime Symmetries and the Indiana University Faculty Research Support Program.


\end{document}